\documentclass[aps,prl,twocolumn,showpacs,superscriptaddress,floatfix]{revtex4-1}

\usepackage{graphicx}
\usepackage{dcolumn}
\usepackage{bm}
\usepackage{amssymb}
\usepackage{microtype}
\usepackage{xfrac}

\begin{document}

\title{Universal Dynamic Magnetism in Yb-Pyrochlores with Disparate Ground States}

\author{A.~M.~Hallas}
\affiliation{Department of Physics and Astronomy, McMaster University, Hamilton, ON, L8S 4M1, Canada}

\author{J.~Gaudet}
\affiliation{Department of Physics and Astronomy, McMaster University, Hamilton, ON, L8S 4M1, Canada}

\author{N.~P.~Butch}
\affiliation{Center for Neutron Research, National Institute of Standards and Technology, MS 6100 Gaithersburg, Maryland 20899, USA}

\author{M.~Tachibana}
\affiliation{National Institute for Materials Science, 1-1 Namiki, Tsukuba 305-0044, Ibaraki, Japan}

\author{R.~S.~Freitas}
\affiliation{Instituto de F\'{i}sica, Universidade de S\~{a}o Paulo, S\~{a}o Paulo 05315-970, SP, Brazil}

\author{G.~M.~Luke}
\affiliation{Department of Physics and Astronomy, McMaster University, Hamilton, ON, L8S 4M1, Canada}
\affiliation{Canadian Institute for Advanced Research, 180 Dundas St. W., Toronto, ON, M5G 1Z7, Canada}

\author{C.~R.~Wiebe}
\affiliation{Department of Physics and Astronomy, McMaster University, Hamilton, ON, L8S 4M1, Canada}
\affiliation{Canadian Institute for Advanced Research, 180 Dundas St. W., Toronto, ON, M5G 1Z7, Canada}
\affiliation{Department of Chemistry, University of Winnipeg, Winnipeg, MB, R3B 2E9 Canada}

\author{B.~D.~Gaulin}
\affiliation{Department of Physics and Astronomy, McMaster University, Hamilton, ON, L8S 4M1, Canada}
\affiliation{Canadian Institute for Advanced Research, 180 Dundas St. W., Toronto, ON, M5G 1Z7, Canada}
\affiliation{Brockhouse Institute for Materials Research, Hamilton, ON L8S 4M1 Canada}

\date{\today}

\begin{abstract}
The ytterbium pyrochlore magnets, Yb$_2B_2$O$_7$ ($B$ = Sn, Ti, Ge) are well described by $S_{\text{eff}} = 1/2$ quantum spins decorating a network of corner-sharing tetrahedra and interacting via anisotropic exchange. Structurally, only the non-magnetic $B$-site cation, and hence, primarily the lattice parameter, is changing across the series. Nonetheless, a range of magnetic behaviors are observed: the low temperature magnetism in Yb$_2$Ti$_2$O$_7$ and Yb$_2$Sn$_2$O$_7$ has ferromagnetic character, while Yb$_2$Ge$_2$O$_7$ displays an antiferromagnetically ordered N\'eel state at low temperatures. While the static properties of the ytterbium pyrochlores are distinct, inelastic neutron scattering measurements reveal a common character to their exotic spin dynamics. All three ytterbium pyrochlores show a gapless continuum of spin excitations, resembling over-damped ferromagnetic spin waves at low $Q$. Furthermore, the specific heat of the series also follows a common form, with a broad, high-temperature anomaly followed by a sharp low-temperature anomaly at $T_C$ or $T_N$. The novel spin dynamics we report correlate strongly with the broad specific heat anomaly \emph{only}, remaining unchanged across the sharp anomaly. This result suggests that the primary order parameter in the ytterbium pyrochlores associated with the sharp anomaly is ``hidden" and not simple magnetic dipole order.
\end{abstract}


\maketitle

The pyrochlores, with chemical composition $A_2B_2$O$_7$, are exemplary realizations of systems that often exhibit strong geometric magnetic frustration when either the $A$ or $B$ site is occupied by a magnetic ion. Recently, the pyrochlore Yb$_2$Ti$_2$O$_7$ has attracted significant interest owing to its rich low temperature physics. Within Yb$_2$Ti$_2$O$_7$, magnetic Yb$^{3+}$ sits on the $A$ site and non-magnetic Ti$^{4+}$ resides on the $B$ site, each forming a network of corner-sharing tetrahedra. The ability to grow large, high-quality single crystals of Yb$_2$Ti$_2$O$_7$ has made it an ideal system for investigation of its microscopic Hamiltonian \cite{PhysRevLett.108.037202,PhysRevB.87.205130,PhysRevX.1.021002,PhysRevLett.106.187202,PhysRevB.92.064425}. These studies suggest that anisotropic exchange is the dominant driver for its low temperature physics. The resulting microscopic Hamiltonian is also the basis for proposals that Yb$_2$Ti$_2$O$_7$ could realize a quantum spin ice state at low temperatures \cite{PhysRevLett.109.097205,YbTiONatComm,PhysRevLett.108.247210,PhysRevB.92.064425,0034-4885-77-5-056501,PhysRevB.84.174442}. 

The low temperature phase behavior in real specimens of Yb$_2$Ti$_2$O$_7$ is complex and subject to strong sample dependencies. The specific heat of polycrystalline samples synthesized by conventional solid state techniques contains a sharp anomaly \cite{PhysRevLett.88.077204}. Meanwhile, single crystal samples grown using zone melting techniques display broader anomalies with lower transition temperatures, multiple transitions, and sometimes no obvious transition \cite{PhysRevB.84.172408,PhysRevB.86.174424,PhysRevB.88.134428}. Furthermore, a host of magnetometry, neutron scattering and muon spin relaxation measurements have shown evidence for an ordered ferromagnetic state below $T_C=265$~mK in some samples of Yb$_2$Ti$_2$O$_7$ \cite{PhysRevB.89.224419,PhysRevLett.88.077204,doi:10.1143/JPSJ.72.3014,YbTiONatComm,PhysRevB.93.064406}, but not in others \cite{PhysRevB.70.180404,PhysRevB.84.174442,PhysRevB.88.134428}. This strong sample dependence, especially in single crystal samples, is attributed to quenched disorder at the 1$\%$ level \cite{PhysRevB.86.174424,PhysRevB.84.172408}. Perhaps moving towards a consensus, recent work on stoichiometric powders of Yb$_2$Ti$_2$O$_7$ has revealed long range splayed ferromagnetic correlations in which the moments are canted towards the local $<$111$>$ direction \cite{PhysRevB.93.064406}. There is, however, a long-standing mystery in Yb$_2$Ti$_2$O$_7$, which is not subject to these sample dependencies \cite{PhysRevB.93.064406}: the absence of conventional spin waves below $T_C$ \cite{PhysRevB.84.174442}. 

The interesting physics of Yb$_2$Ti$_2$O$_7$ provided the initial impetus for the study of Yb$_2$Sn$_2$O$_7$, in which the non-magnetic $B$-site has been substituted with Sn$^{4+}$. The larger ionic radius of Sn$^{4+}$ increases the lattice parameter of Yb$_2$Sn$_2$O$_7$ by approximately 3\%. Yb$_2$Sn$_2$O$_7$ undergoes an ordering transition at $T_C=150$~mK to a splayed ferromagnetic state \cite{PhysRevLett.110.127207} in which the moments are canted away from ferromagnetic alignment along $<$100$>$ but in the opposite direction from Yb$_2$Ti$_2$O$_7$. Despite this transition in Yb$_2$Sn$_2$O$_7$ being marked by a relatively sharp anomaly in the specific heat \cite{PhysRevLett.110.127207,PhysRevB.89.024421}, there is an absence of conventional spin wave excitations below $T_C$, as in the case of Yb$_2$Ti$_2$O$_7$ \cite{PhysRevB.87.134408}.


Entering the fray most recently is Yb$_2$Ge$_2$O$_7$, where the $B$-site is now occupied by non-magnetic Ge$^{4+}$. The smaller ionic radius of germanium reduces the lattice parameter by 2\% as compared to Yb$_2$Ti$_2$O$_7$. In the first report on Yb$_2$Ge$_2$O$_7$, non-linear ac susceptibility measurements showed an ordering transition at $T_N=0.57$~K with antiferromagnetic character \cite{PhysRevB.89.064401}. Subsequent neutron diffraction measurements have identified the antiferromagnetic $k=0$ ordered state in Yb$_2$Ge$_2$O$_7$ as belonging to the $\Gamma_5$ manifold \cite{PhysRevB.92.140407,Hallas_YGO_PRB}. In the $\Gamma_5$ ordered state, the four moments on each tetrahedron are oriented perpendicular to the local $<$111$>$ axes, a markedly different state than those observed in the other ytterbium pyrochlores. 

\begin{figure}[tbp]
\linespread{1}
\par
\includegraphics[width=3.2in]{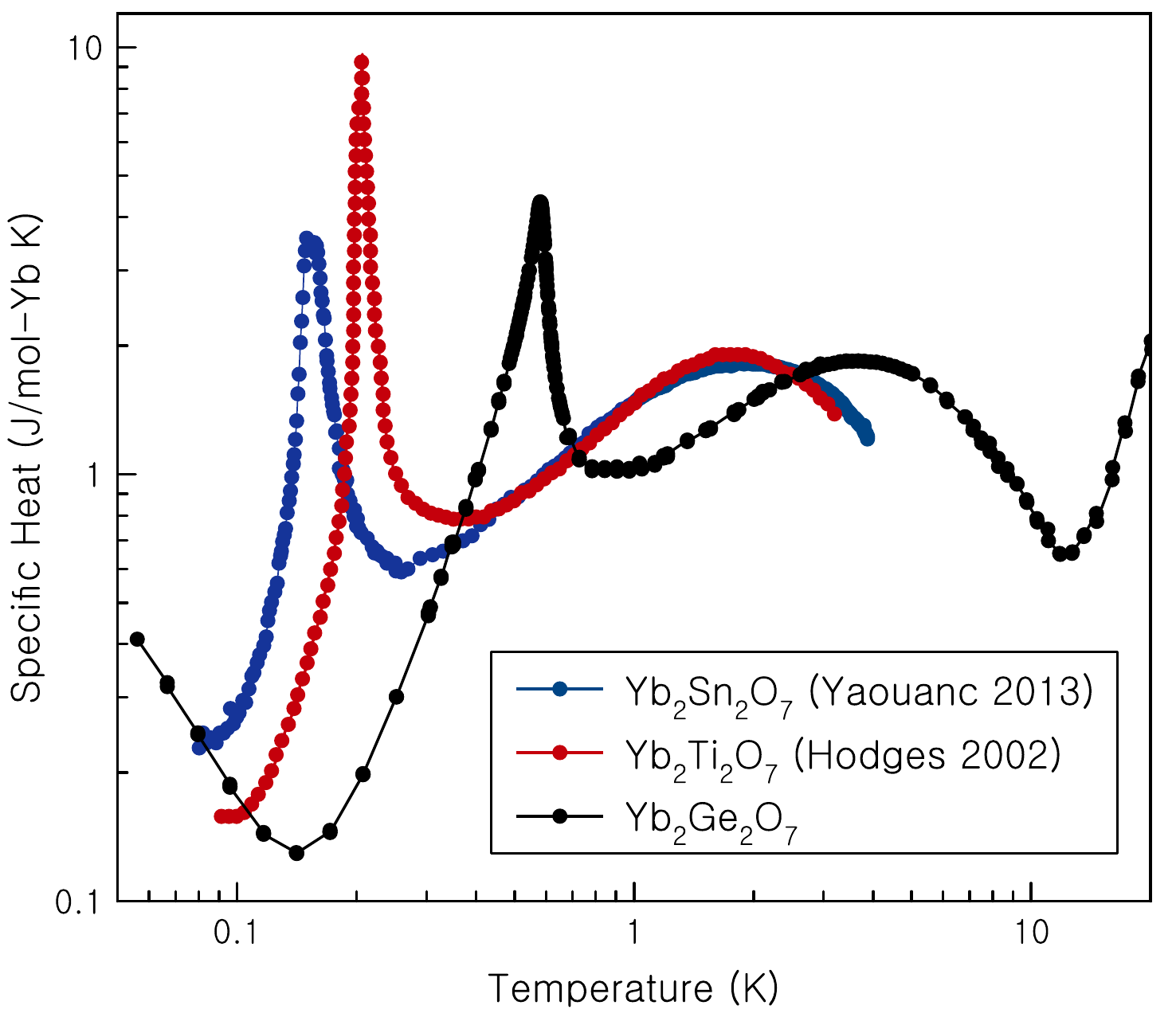}
\par
\caption{The specific heat of the ytterbium pyrochlores, Yb$_2B_2$O$_7$ ($B$ = Ge, Ti, and Sn), where in each case a broad high-temperature anomaly precedes a sharp low-temperature anomaly. The upturn at the lowest temperatures is due to a nuclear Schottky anomaly.}
\label{Fig1}
\end{figure}

There are a number of distinguishing magnetic properties for these three ytterbium pyrochlores, Yb$_2B_2$O$_7$ ($B$ = Sn, Ti, Ge), as described above. However, there are commonalities as well. Neutron spectroscopy of Yb$_2$Ti$_2$O$_7$ and Yb$_2$Ge$_2$O$_7$ has revealed that their crystal electric field schemes are largely the same \cite{Hallas_YGO_PRB,PhysRevB.92.134420}. Due to their well-separated ground state doublets, both systems are described in terms of $S_{\text{eff}}=\sfrac{1}{2}$ quantum spins with local XY anisotropy. Similar crystal field phenomenology is expected for Yb$_2$Sn$_2$O$_7$. Perhaps most striking, however, is the similarity in the low temperature specific heat of these three materials (Fig.~\ref{Fig1}). In each case, there is a broad specific heat anomaly at high-temperature that precedes a sharp low-temperature anomaly \cite{PhysRevLett.88.077204,PhysRevLett.110.127207}. This sharp, $\lambda$-like specific heat anomaly heralds the onset of long range magnetic correlations as detected by neutron diffraction in Yb$_2$Sn$_2$O$_7$ \cite{PhysRevLett.110.127207} and Yb$_2$Ge$_2$O$_7$ \cite{Hallas_YGO_PRB}; long range correlations are also found in Yb$_2$Ti$_2$O$_7$ but with an onset temperature higher than $T_C$ \cite{PhysRevB.93.064406}. Furthermore, in each case, the total integrated magnetic entropy is consistent with $R\ln{(2)}$, the expected value for a well-separated ground state doublet.

In this letter, we present a new detailed study of the dynamic properties of Yb$_2$Ge$_2$O$_7$ via inelastic neutron scattering. We then compare these results to those for Yb$_2$Ti$_2$O$_7$ and Yb$_2$Sn$_2$O$_7$ and discover a ubiquitous character to their exotic magnetic excitations. In each case, this excitation spectrum correlates strongly with {\it only} the broad, high temperature specific heat anomaly. This is a striking result, as these three materials do not share a common ordered magnetic dipole state. Thus, it may be the case that the primary order in the ytterbium pyrochlores associated with the sharp, low temperature anomaly is ``hidden", to use an analogy to the enigmatic hidden order state displayed by URu$_2$Si$_2$ \cite{URu2Si2}.


\begin{figure}[tbp]
\linespread{1}
\par
\includegraphics[width=3.2in]{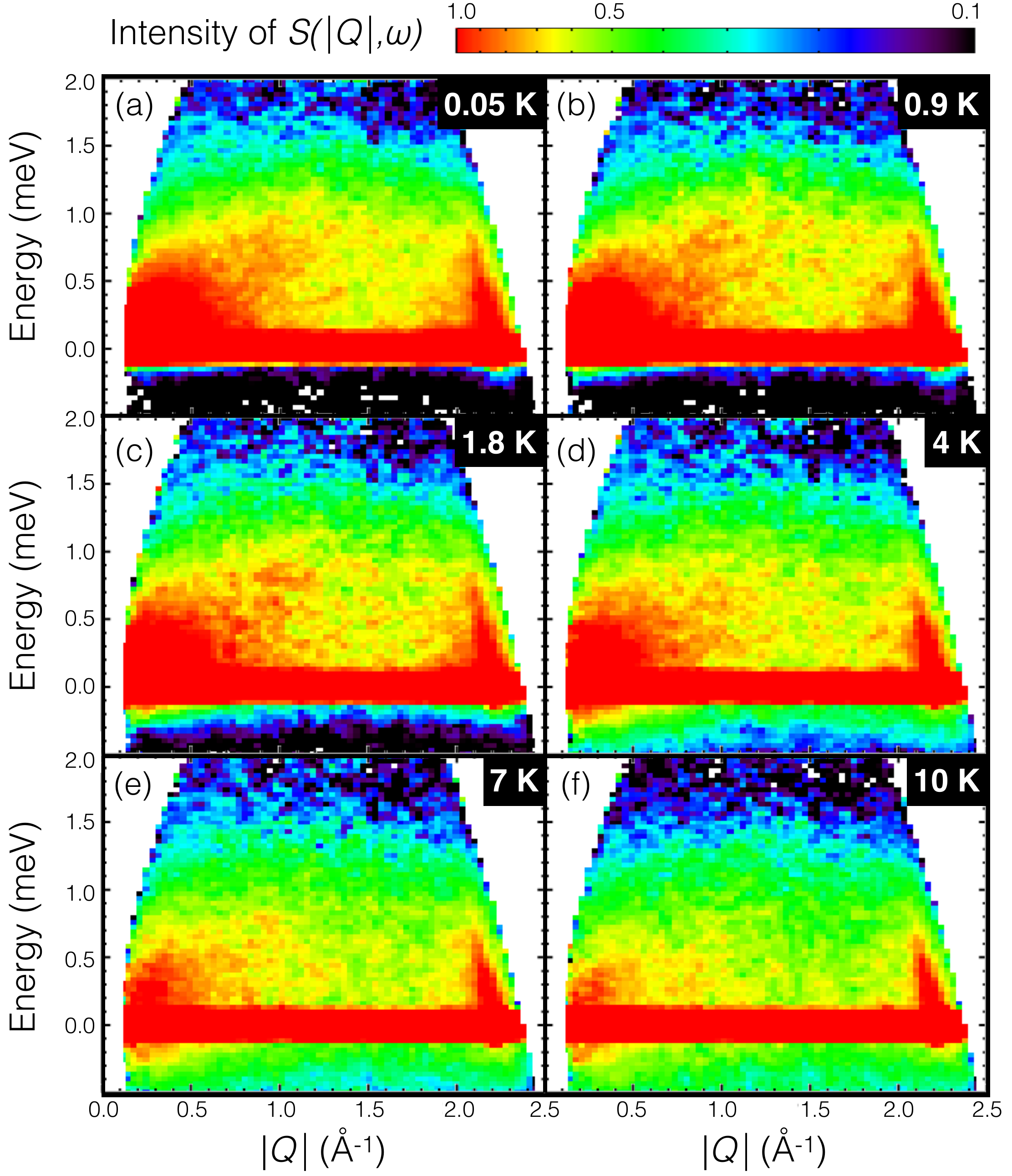}
\par
\caption{Temperature dependence of the inelastic scattering, $S(Q, \omega)$, of Yb$_2$Ge$_2$O$_7$ measured at (a) 0.05~K (b) 0.9~K (c) 1.8~K (d) 4~K (e) 7~K and (f) 10~K. In each case, the background was subtracted using an empty can measurement. All six data sets have been normalized by counting time and scaled over an identical logarithmic intensity range.}
\label{Fig2}
\end{figure}

\begin{figure*}[htbp]
\linespread{1}
\par
\includegraphics[width=7in]{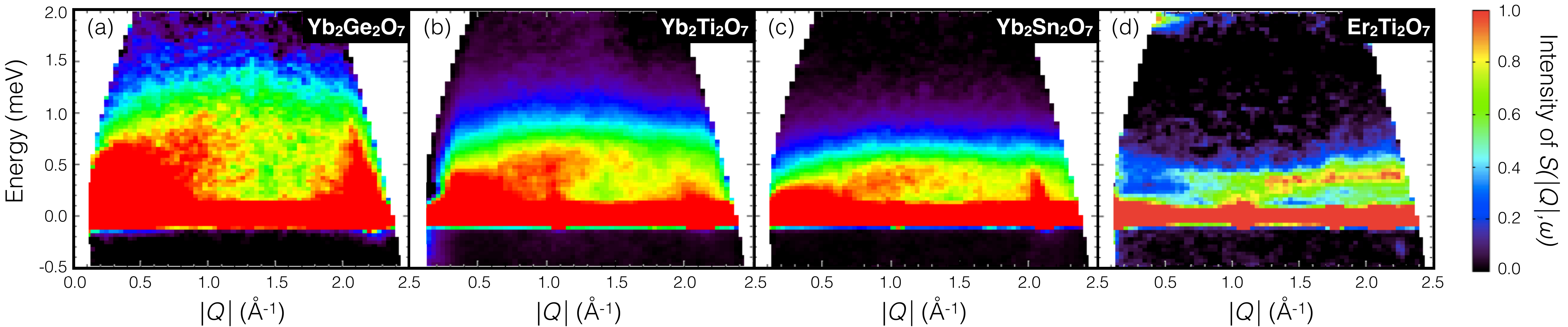}
\par
\caption{Inelastic scattering, $S(Q, \omega)$, of (a) Yb$_2$Ge$_2$O$_7$ (b) Yb$_2$Ti$_2$O$_7$ (c) Yb$_2$Sn$_2$O$_7$ and (d) Er$_2$Ti$_2$O$_7$. For each sample the measurement was taken at a temperatures of 100~mK or lower and an empty can background has been subtracted. In the cases of the ytterbium pyrochlores, Yb$_2B_2$O$_7$, the measurements were performed on powder samples. The measurement on Er$_2$Ti$_2$O$_7$, which was taken on a single crystal sample, has been powder averaged over the $(HHL)$ scattering plane. The intensity ranges have been selected such that the inelastic scattering at $Q$~=~1.2~\AA$^{-1}$ saturates the scale.}
\label{Fig3}
\end{figure*}


The experimental details are given in the Supplemental Material. Figure~\ref{Fig2} shows the inelastic neutron scattering spectra, $S(Q, \omega)$, for Yb$_2$Ge$_2$O$_7$ collected at temperatures ranging from 0.05~K to 10~K, above and below both specific heat anomalies. At all temperatures, we observe an intense phonon mode originating from the $(222)$ Bragg position at 2.3~\AA$^{-1}$. However, we attribute all other inelastic signal to a magnetic origin. At 50~mK the inelastic scattering in Yb$_2$Ge$_2$O$_7$ forms a continuum of spin excitations that is gapless within the resolution of our measurement (0.09~meV). These diffusive excitations are maximally intense approaching $Q=0$ and resembles over-damped ferromagnetic spin waves ($E \propto Q^2$) at the lowest wave vectors. 
Integration of the elastic scattering reveals the formation of magnetic Bragg reflections between 50~mK and 900~mK that give an ordered moment consistent with previous diffraction measurements (See Supplemental Material) \cite{Hallas_YGO_PRB}. Yet, there is little or no temperature dependence to the spin excitation spectrum across $T_N$, between 50~mK and 1.8~K. At 4~K, there is a small decrease in the inelastic intensity, particularly at low~$Q$, while at 7~K and 10~K the scattering is significantly reduced at all values of $Q$. Furthermore the signal is shifted lower in energy, softening towards the elastic line. These spin excitations in Yb$_2$Ge$_2$O$_7$ are clearly exotic in two respects: (i) firstly, they do not correlate in temperature with the sharp specific heat anomaly at $T_N = 0.57$~K, but instead with the broad specific heat anomaly centered at 4~K (ii) secondly, these are far removed from resolution-limited spin wave excitations that would be expected for \emph{any} ordered state.

It is illuminating to compare our lowest temperature inelastic neutron scattering measurements on Yb$_2$Ge$_2$O$_7$ to those of its fellow XY pyrochlores. Neutron diffraction measurements have found that the ordered state in Yb$_2$Ge$_2$O$_7$ belongs to the $k=0$, $\Gamma_5$ irreducible representation, the same ground state manifold to which Er$_2$Ti$_2$O$_7$ belongs \cite{PhysRevB.68.020401}. While $\Gamma_5$ does contain two basis vectors, $\psi_2$ and $\psi_3$, the powder diffraction patterns for these two states are identical and thus, cannot be distinguished in polycrystalline Yb$_2$Ge$_2$O$_7$. Regardless, the spin wave spectra arising from these two states contain only subtle differences \cite{PhysRevLett.109.167201}. Thus, whether Yb$_2$Ge$_2$O$_7$ orders into $\psi_2$, $\psi_3$, or a linear combination of the two, we would expect the spin wave spectra to strongly resemble that of Er$_2$Ti$_2$O$_7$. However, comparing the inelastic scattering profiles of Yb$_2$Ge$_2$O$_7$ (Fig.~\ref{Fig3}(a)) and Er$_2$Ti$_2$O$_7$ (Fig.~\ref{Fig3}(d)), both in their ordered states, it is clear that this is not the case. In Er$_2$Ti$_2$O$_7$, the sharp, Goldstone-like spin waves arise from the (111) Bragg peak at 1.1~\AA$^{-1}$ with an absence of scattering around $Q=0$, typical for a conventional antiferromagnet. Conversely, the excitations in Yb$_2$Ge$_2$O$_7$ are extremely diffusive and maximally intense approaching $Q=0$, typical of a disordered ferromagnet. 

We instead find that the excitations in Yb$_2$Ge$_2$O$_7$ strongly resemble those of Yb$_2$Ti$_2$O$_7$ (Fig.~\ref{Fig3}(b)) and Yb$_2$Sn$_2$O$_7$ (Fig.~\ref{Fig3}(c)). This is a remarkable result as these three ytterbium pyrochlores \emph{do not} share an ordered state. In fact, Yb$_2$Ge$_2$O$_7$ orders antiferromagnetically, while Yb$_2$Sn$_2$O$_7$ order ferromagnetically and Yb$_2$Ti$_2$O$_7$ shows related ferromagnetism. The diffusive spin excitations, centered at $Q=0$, share a common form in each of these materials, and appear to simply scale in energy across the series. Although this excitation spectrum is ferromagnetic in appearance, it is gapless at all wavevectors and does not resemble the gapped spin wave spectrum expected to arise due to anisotropic exchange within a splayed ferromagnetic ordered state \cite{PhysRevB.93.064406}. Furthermore, for all three materials, the excitations evolve in temperature only in the range of the broad specific heat anomaly and do not obviously acknowledge the presence of the sharp low temperature specific heat anomaly \cite{PhysRevB.93.064406,PhysRevB.87.134408}. This is quite distinct from the low temperature properties of Er$_2$Ti$_2$O$_7$, where a single specific heat anomaly marks its transition to the $\psi_2$ antiferromagnetic state, and this directly correlates with the evolution of well-defined spin wave excitations \cite{PhysRevB.68.020401,Blote1969549,PhysRevLett.101.147205} and a spin wave gap \cite{PhysRevLett.112.057201}. 

\begin{figure}[tbp]
\linespread{1}
\par
\includegraphics[width=3.2in]{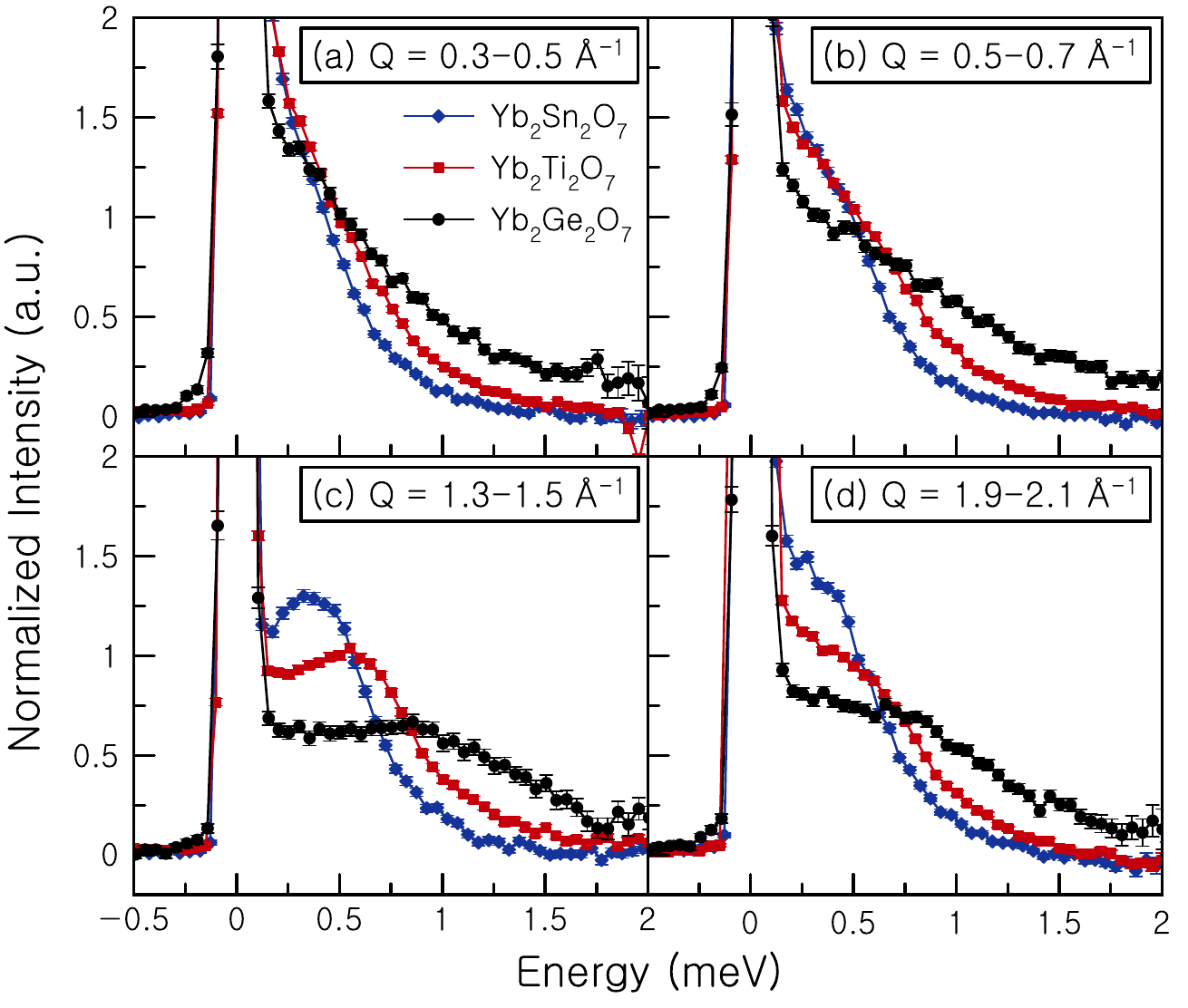}
\par
\caption{Integrated scattering intensity of Yb$_2B_2$O$_7$ ($B$ = Ge, Ti, and Sn) as a function of energy transfer over the $Q$ intervals (a) $0.3-0.5$ \AA$^{-1}$ (b) $0.5-0.7$ \AA$^{-1}$ (c) $1.3-1.5$ \AA$^{-1}$ and (d) $1.9-2.1$ \AA$^{-1}$. The intensities have been normalized by scaling the simulated relative intensities on the largest nuclear Bragg reflection, (222).}
\label{Fig4}
\end{figure}

In order to further compare the inelastic scattering spectra of Yb$_2B_2$O$_7$ ($B$ = Ge, Ti, and Sn) we have taken several integrated cuts through the normalized data over selected $Q$-ranges. For each interval, the same general $Q$ dependence is observed in the three materials. At the lowest values of $Q$ the intensity is primarily quasielastic but at larger values of $Q$, the spectral weight begins to shift to higher energies. The bandwidth of the spin excitations is smallest in Yb$_2$Sn$_2$O$_7$ and the most extended in Yb$_2$Ge$_2$O$_7$. The top of the bandwidth for Yb$_2B_2$O$_7$ ($B$ = Ge, Ti, and Sn), which occurs in the interval between $Q = [1.3,1.5]$~\AA$^{-1}$, are 0.9~meV, 0.69~meV and 0.48~meV, respectively (Fig.~\ref{Fig4}(c)). This continuum of scattering appears to be arcing towards the (222) Bragg reflection, and accordingly over this $Q$-interval the intensity shifts back towards the elastic line. It is, however, important to note that the (222) nuclear Bragg reflection is intense for each of these materials and thus, phonon scattering partially obscures this region. 

\begin{figure}[tbp]
\linespread{1}
\par
\includegraphics[width=3.2in]{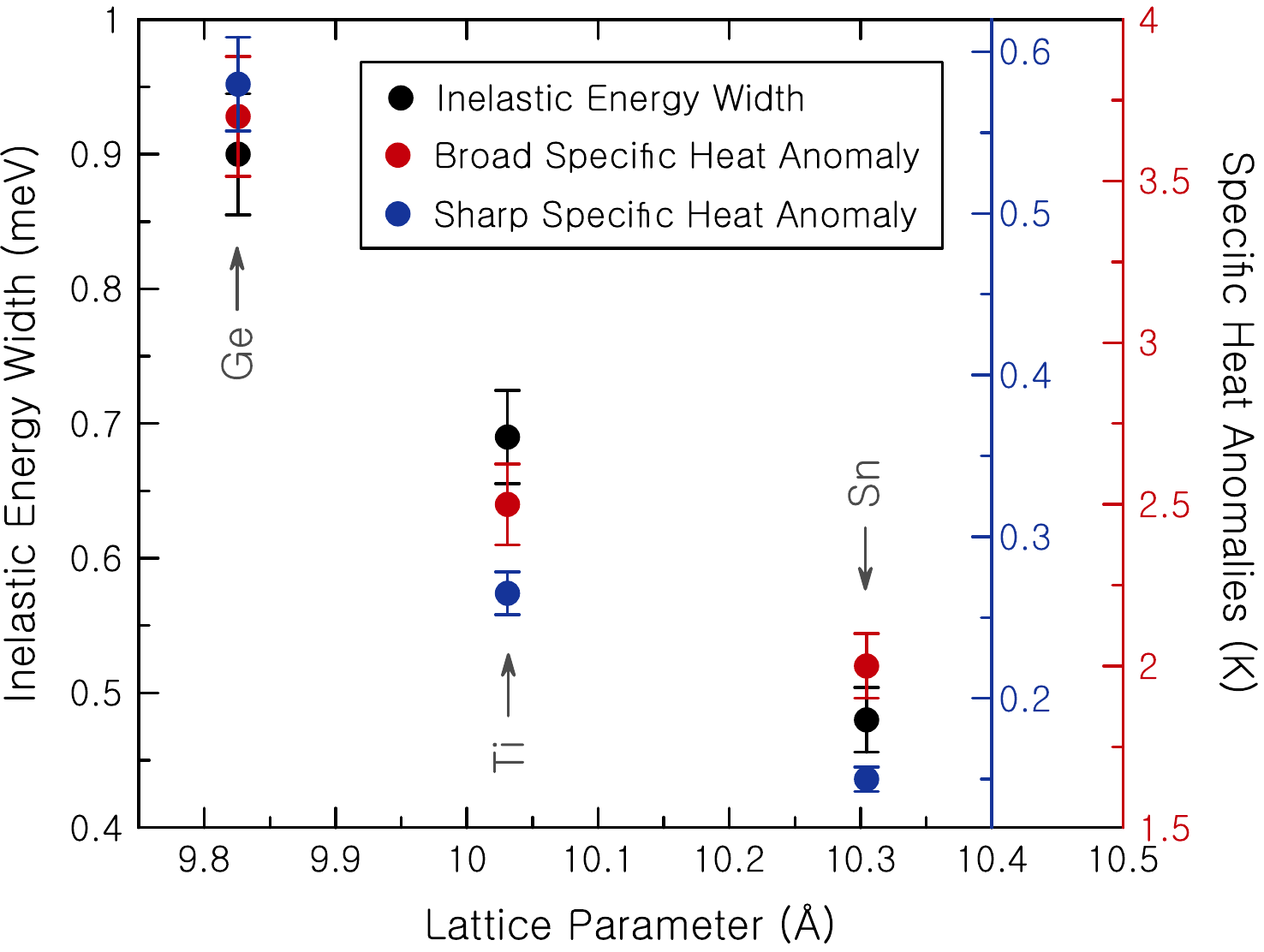}
\par
\caption{The relationship between lattice parameter and magnetism temperature scales in Yb$_2B_2$O$_7$ ($B$ = Ge, Ti, and Sn). The temperatures at which the heat capacity anomalies appear, as well as the energy bandwidth of the inelastic scattering scales linearly with the lattice parameter.
}
\label{Fig5}
 \end{figure}

In Figure~\ref{Fig5} we plot the energy bandwidth for the spin excitations, as defined above, as well as the temperatures for the two specific heat anomalies in Yb$_2B_2$O$_7$ ($B$ = Ge, Ti, and Sn) as a function of their lattice parameter. We find a remarkably linear correspondence between the bandwidth and the lattice parameter, as well as the temperature of the broad heat capacity anomaly which we attribute to the formation of these excitations. It is interesting to note that even the temperature of the sharp heat capacity anomaly tracks quite well with this dependence, despite having no apparent role in the observed continuum of diffusive magnetic scattering. It is also worth noting that since we observe magnetic Bragg reflections for Yb$_2$Ge$_2$O$_7$ in a $\Gamma_5$ ordered state, there should be some signature of this in the inelastic signal as well. We conjecture that these spin waves are in fact present, but are overwhelmed by the intensity of the continuum of magnetic scattering. While only approximately 0.3(1) $\mu_{\text{B}}$ of the ytterbium moment is involved in this ordered state of Yb$_2$Ge$_2$O$_7$ \cite{Hallas_YGO_PRB}, it appears that the majority of the ytterbium moment is contributing to the ferromagnetic-like continuum of scattering.

The universality in the nature of the spin excitation spectrum in Yb$_2B_2$O$_7$ ($B$ = Ge, Ti, and Sn), as well as in the form of the specific heat, strongly suggests that the prevailing order is {\it not} driven primarily by magnetic dipole correlations. The spin excitation spectrum itself, as measured by inelastic neutron scattering, is a measure of magnetic dipole fluctuations on an appropriate energy scale. Thus, there are two key observations that suggest magnetic dipole order is not the primary order parameter in Yb$_2B_2$O$_7$. Firstly, $S(Q, \omega)$ shows little or no temperature dependence on a scale related to the ordering temperature in any of these materials. Secondly, the form of $S(Q, \omega)$ is the same across the series despite different magnetic dipole ordered ground states. A similar scenario appears to be realized in a different pyrochlore magnet, Tb$_{2+x}$Ti$_{2-x}$O$_{7+y}$ with $x=0.005$, in which the low temperature thermodynamic phase transition doesn't obviously correlate with its magnetic neutron order parameter \cite{2015arXiv151100733K}. In Tb$_{2+x}$Ti$_{2-x}$O$_{7+y}$ with $x=0.005$ it has been speculated that this order originates from multipolar correlations. However, the well-separated ground state doublet in Yb$_2B_2$O$_7$ precludes the possibility of a multipolar scenario. Hence, we draw an association with ``hidden" order.

Finally, it is worth emphasizing that it is not only the shared form of these excitations that is exotic. The spin excitations are highly exotic in their own right - a gapless continuum of scattering, unrelated to the magnetic dipole ordered state of each material. Especially in Yb$_2$Ge$_2$O$_7$, where the ordered state is a $\Gamma_5$ antiferromagnet, it is clear that these excitations, which resemble over-damped ferromagnetic spin waves, have little relation to the dipole ordered state. Furthermore, while Yb$_2$Ti$_2$O$_7$ is renowned for sample dependence and sensitivity to weak disorder \cite{PhysRevB.86.174424,PhysRevB.84.172408}, these dynamic properties stand in stark contrast; the spin excitations are robust, appearing in both powder and single crystalline samples, as well as in non-stoichiometric samples \cite{PhysRevB.93.064406}. As we associate these excitations with the broad specific heat anomaly, which accounts for approximately 80\% of the $R\ln{(2)}$ entropy, our results should prompt a refocusing of the theoretical studies which attempt to understand the magnetism of the Yb-pyrochlores. It is apparent that these unconventional excitations hold the key to a comprehensive understanding of this exotic family of quantum magnets.

To conclude, new neutron scattering measurements on Yb$_2$Ge$_2$O$_7$ reveal the same exotic form to the spin dynamical properties, $S(Q, \omega)$, as have been observed in Yb$_2$Ti$_2$O$_7$ and Yb$_2$Sn$_2$O$_7$. This is despite the fact that, below $T_N=0.57$~K, the magnetic dipole ordered phase in Yb$_2$Ge$_2$O$_7$ is a $\Gamma_5$ antiferromagnet, rather than splayed ferromagnetism, as in Yb$_2$Ti$_2$O$_7$ and Yb$_2$Sn$_2$O$_7$. In all three of the ytterbium pyrochlores, $S(Q, \omega)$ evolves on a temperature scale set {\it only} by the broad, high temperature specific heat anomaly. This broad, high temperature feature in the specific heat carries with it approximately 80$\%$ of the $R\ln{(2)}$ entropy in Yb$_2$Ge$_2$O$_7$, and is associated with most of the dipole moment and spectral weight probed by neutron scattering. This strongly suggests that the topical ground states in the Yb$_2B_2$O$_7$ ($B$ = Ge, Ti and Sn) series are perhaps more exotic than previously thought, with a dominant ``hidden" order parameter at the lowest temperatures.

\begin{acknowledgments}
We acknowledge useful conversations with M.J.P. Gingras and L. Balents. We thank H.D. Zhou, K.A. Ross, and E. Kermarrec for making their data available for this work. A.M.H. acknowledges support from the Vanier Canada Graduate Scholarship Program and thanks the National Institute for Materials Science (NIMS) for their hospitality and support through the NIMS Internship Program. This work was supported by the Natural Sciences and Engineering Research Council of Canada and the Canada Foundation for Innovation. C.R.W. acknowledges support through the Canada Research Chairs Program (Tier II). Work at the NIST Center for Neutron Research is supported in part by the National Science Foundation under Agreement No. DMR-0944772. R.S.F. acknowledges support from CNPq (Grant No. 400278/2012­0).
\end{acknowledgments}

\bibliography{YGO_Inelastic}

\end{document}